%Paper: hep-ph/9404277
%From: NARISON@crnvma.cern.ch
%Date: Mon, 18 Apr 94 15:58:06 SET

%*******************************************************************
%FONTS

\magnification 1200
\font\bigbf= cmbx10 scaled \magstep1

%*******************************************************************
%FOOTNOTES - insert command \eightpoint to reduce

\newskip\ttglue

\font\eightrm=cmr8
\font\eighti=cmmi8
\font\eightsy=cmsy8
\font\eightbf=cmbx8
\font\eighttt=cmtt8
\font\eightsl=cmsl8
\font\eightit=cmti8
\font\sixrm=cmr6
\font\sixbf=cmbx6
\font\sixi=cmmi6
\font\sixsy=cmsy6

\def \eightpoint{\def\rm{\fam0\eightrm}% switch to 8-point type
\textfont0=\eightrm \scriptfont0=\sixrm \scriptscriptfont0=\fiverm
\textfont1=\eighti \scriptfont1=\sixi   \scriptscriptfont1=\fivei
\textfont2=\eightsy \scriptfont2=\sixsy   \scriptscriptfont2=\fivesy
\textfont3=\tenex \scriptfont3=\tenex   \scriptscriptfont3=\tenex
\textfont\itfam=\eightit  \def\it{\fam\itfam\eightit}%
\textfont\slfam=\eightsl  \def\sl{\fam\slfam\eightsl}%
\textfont\ttfam=\eighttt  \def\tt{\fam\ttfam\eighttt}%
\textfont\bffam=\eightbf  \scriptfont\bffam=\sixbf
 \scriptscriptfont\bffam=\fivebf  \def\bf{\fam\bffam\eightbf}%
\tt \ttglue=.5em plus.25em minus.15em
\setbox\strutbox=\hbox{\vrule height7pt depth2pt width0pt}%
\normalbaselineskip=9pt
\let\sc=\sixrm  \let\big=\eightbig  \normalbaselines\rm
}
%*******************************************************************
%GREEK LETTERS
\def\a{\alpha}
\def\b{\beta}
\def\c{\gamma}
\def\d{\delta}
\def\e{\epsilon}
\def\h{\eta}

\def\m{\mu}
\def\n{\nu}

\def\p{\pi}
\def\r{\rho}

\def\t{\tau}

\def\C{\Gamma}
\def\D{\Delta}
\def\L{\Lambda}

%*****************************************************************
%OTHER MACROS
\def\pl{\partial}
\def\ra{\rightarrow}
\def\vev{\langle \phi \rangle}
\def\hp{\eta^{\prime}}

\def\OO{{\cal O}}

\def\MV{{\rm MeV}}
\def\GV{{\rm GeV}}

\def\tr{{\rm tr}}
\def\el{{\rm e}}
\def\bigQ{\mathrel{\mathop\sim_{Q^2\to\infty}}}
\def\tc{{\tilde \gamma}}

%*****************************************************************
%TITLE PAGE
{\vsize 22cm
{\nopagenumbers

%\vskip0.5cm
\line{\hfil PM/94-14         }
\line{\hfil SWAT-94/12          }
\line{\hfil CERN-TH.7223/94     }

\vskip1cm
\centerline{{{\bigbf TARGET INDEPENDENCE OF THE
EMC-SMC EFFECT       }} }
%\vskip0.5cm
%\centerline{{\bf FROM QCD SUM RULES}}
\vskip1cm
\centerline{\bf S. Narison}
\vskip0.2cm
\centerline{\it TH Division, CERN}
\centerline{\it CH 1211 Geneva 23, Switzerland}
\vskip0.1cm
\centerline{\it and}
\vskip0.1cm
\centerline{\it Laboratoire de Physique Math\'ematique (URA 768), USTL,}
\centerline{\it place E.~Bataillon, 34090 Montpellier Cedex 6, France}
\vskip0.7cm
\centerline{\bf G.M. Shore}
\vskip0.2cm
\centerline{\it Department of Physics, }
\centerline{\it University of Wales, Swansea }
\centerline{\it Swansea, SA2 8PP, U.K. }
\vskip0.7cm
\centerline{\bf G. Veneziano}
\vskip0.2cm
\centerline{\it TH Division, CERN}
\centerline{\it CH 1211 Geneva 23, Switzerland}
\vskip1cm
\noindent{\bf Abstract}
\vskip0.2cm
\noindent An approach to deep inelastic scattering
is described in which the matrix elements arising from the
operator product expansion are factorised into composite operator
propagators and proper vertex functions. In the case of polarised
$\m p$ scattering, the composite operator propagator is identified
with the square root of the QCD topological susceptibility
$\sqrt{\chi^{\prime}(0)}$, while the corresponding proper vertex is
a renormalisation group invariant. We estimate $\chi^{\prime}(0)$
using QCD spectral sum rules and find that it is significantly
suppressed relative to the OZI expectation.
Assuming OZI is a good approximation for the proper vertex,
our predictions, $\int_{0}^{1}dx g_1^p (x;Q^2=10\GV^2)= 0.143
\pm 0.005$ and $G^{(0)}_A \equiv \Delta \Sigma = 0.353 \pm 0.052$,
are in excellent agreement with the new SMC data.
This result, together with one confirming the validity of the OZI
rule in the $\hp$ radiative decay, supports our earlier conjecture
that the suppression in the flavour singlet component of
the first moment of $g_1^p$ observed by the EMC-SMC collaboration is
a target-independent feature of QCD related to the $U(1)$
anomaly and is not a property of the proton structure.
As a corollary, we extract the magnitude of higher twist effects
from the neutron and Bjorken sum rules.

\vskip0.7cm
\line{PM/94-14       \hfil}
\line{SWAT-94/12  \hfil}
\line{CERN-TH.7223/94 \hfil}
\line{April 1994  \hfil}

\vfill\eject } }

\pageno = 1

\noindent{\bf 1. Introduction}
\vskip0.5cm
The discovery by the EMC collaboration[1] (see also [2]) of an
unexpected suppression of the first moment of the polarised proton
structure function $g_1^p$ has provoked an extensive discussion
of the parton model interpretation of QCD in deep inelastic
scattering processes involving the axial $U(1)$ anomaly. (For
reviews, see refs.[3,4].) While it has so far proved possible
with careful redefinitions and interpretations[5] to preserve
the essence of the parton model description, it is becoming clear
that these processes involve subtle field theoretic properties of QCD
which lead beyond both the original and QCD-improved parton
approximation. In this paper, we develop an alternative approach
to deep inelastic scattering emphasising field theoretic concepts
such as the operator product expansion (OPE), composite operator
Green functions and proper vertices. This clarifies some of the
difficulties encountered in the parton description and gives a
new insight into the underlying reason for the EMC result.
In particular, our analysis strongly suggests that the observed
suppression of the first moment of $g_1^p$ is a generic QCD effect
related to the anomaly and is actually {\it independent} of the target.
Rather than revealing a special property of the proton structure,
the EMC result reflects an anomalously small value of the first
moment of the QCD topological susceptibility[6,7].

The essential features of this method are easily described for a
general deep inelastic scattering process. The hadronic part of the
scattering amplitude is given by the imaginary part of the
two-current matrix element $\langle N| J_\m (q)~J_\n(-q) |N\rangle$
illustrated in Fig.~1, where $J_\m$ is the current coupling to the
exchanged hard photon (or electroweak vector boson) and $|N\rangle$
denotes the target. The OPE is used to expand the large $Q^2$ limit
of the product of currents as a sum of Wilson coefficients
$C_i(Q^2)$ times renormalised composite operators $\OO_i$ as
follows (suppressing Lorentz indices),
$$
J(q)~J(-q)~~\bigQ ~~\sum_i~C_i(Q^2)~\OO_i(0)
.\eqno(1.1)
$$
The dominant contributions to the amplitude arise from the operators
$\OO_i$ of lowest twist. Within this set of lowest twist operators,
those of spin $n$ contribute to the $n$th moment of the structure
functions, i.e.
$$
\int_0^1 dx ~x^{n-1} F(x,Q^2) ~~=~~ \sum_i~C_i^n(Q^2)~\langle N|
\OO_i^n(0) |N\rangle
.\eqno(1.2)
$$

The Wilson coefficients are calculable in QCD perturbation theory,
so the problem reduces to evaluating the target matrix elements of the
corresponding operators. We now introduce appropriately defined
proper vertices $\C_{\tilde\OO NN}$, which are chosen to be 1PI
with respect to a physically motivated basis set $\tilde\OO_k$
of renormalised composite operators. The matrix elements are then
decomposed into products of these vertices with zero-momentum
composite operator propagators as follows,
$$
\langle N| \OO_i(0) |N\rangle ~~=~~ \sum_k~\langle 0|\OO_i(0)~
\tilde\OO_k(0) |0\rangle~ \C_{\tilde\OO_k NN}
.\eqno(1.3)
$$
\vfill\eject
{\vsize20cm
This is illustrated in Fig.~2. In essence, what we have done is to
split the whole amplitude into the product of a ``hot'' (high
momentum) part described by QCD perturbation theory, a ``cold''
part described by a (non-perturbative) composite operator
propagator and finally a target-dependent proper vertex.

All the target dependence is contained in the vertex function
$\C_{\tilde\OO NN}$. However, these are not unique -- they depend
on the choice of the basis $\tilde\OO_k$ of composite operators.
This choice is made on physical grounds based on the relevant
degrees of freedom, the aim being to parametrise the amplitude
in terms of a minimal, but sufficient, set of vertex functions.
A good choice can often lead to an almost direct correspondence
between the proper vertices and physical couplings such as, e.g.,
the pion-nucleon coupling $g_{\p NN}$. In particular, it will be
wise to use, whenever possible, RG-invariant proper vertices.

Despite being non-perturbative, we can frequently evaluate the
composite operator Green functions using a combination of
exact Ward identities and dynamical approximations (see sects.~2 and 3).
On the other hand, because of the target dependence, the proper
vertices are not readily calculable from first principles in QCD,
so we are in general left with a parametrisation of the amplitude in
terms of a (hopefully small) set of unknown vertices.
These play the r\^ole of the non-perturbative (i.e. primordial
or not-yet-evolved) parton distributions in the usual
treatment. Just as for parton distributions, many different QCD
processes can be related through parametrisation with the same
set of vertex functions.

Now compare this approach with the parton model.
In the original parton model, the amplitude is approximated
by Fig.~3, describing the scattering of a large $Q^2$ photon
with a parton in the target nucleon. This picture is already
sufficient to reveal Bjorken scaling. It may be improved in the
context of QCD by including gluonic corrections, exactly as in the OPE,
as shown in Fig.~4. These give the logarithmic scaling violations
characteristic of perturbative QCD. The total amplitude is therefore
factorised into a perturbative scattering amplitude for the hard photon
with a parton (quark or gluon) and a parton distribution function giving
the probability of finding a particular parton with given fraction $x$ of
the target momentum.

The question of whether the full QCD amplitude can be given a natural
parton interpretation depends on the composite operators $\OO_i$
in the Wilson expansion. For example, if the lowest twist operator
for a given process is multilinear in the elementary quark and gluon
fields rather than simply quadratic then the diagram of Fig.~4 is
not appropriate and the process can only be described in terms of
multi-parton distributions[8]. A more subtle problem arises when the
operators $\OO_i$ are non-trivially renormalised and mix with other
composite operators under renormalisation. In this case, the parton
interpretation is preserved by {\it defining} parton distributions
directly in terms of the operator matrix elements (see, e.g., ref.[8]).
This procedure becomes especially delicate[5] in the case of
polarised deep inelastic scattering because of the special
renormalisation properties of the relevant Wilson operator
$J^0_{\m 5R}$ due to the axial $U(1)$ anomaly.

In this paper, rather than attempt to interpret the amplitude for
polarised deep inelastic scattering in terms of specially defined
polarised quark and gluon distributions, we instead focus the
analysis on the composite operator level. By splitting the matrix
elements in the form of eq.(1.3), we can exploit chiral Ward identities
and the renormalisation group to separate out generic features of QCD
manifested in the composite operator propagator from specific
properties of the target. In the next section, we see how this
clarifies the origin of the suppression of the first moment of
$g_1^p$ observed in polarised $\m p$ scattering.

\vfill\eject  }

\noindent{\bf 2. The First Moment Sum Rule for $g_1^p$ }
\vskip0.5cm
Our starting point is the familiar Ellis-Jaffe[9]
sum rule for the first moment
of the polarised proton structure function $g_1^p$. For
$N_F=3$ and in the $\overline{MS}$ scheme[10], this reads\footnote{$^*$}
{\eightpoint
\noindent We use the NLO and NNLO coefficients given in Ref.~[11].
However, due to our definition (2.5) of the renormalised composite
operators, the radiative corrections of the singlet are different
from the corresponding terms in ref.[11], which uses a different
renormalisation of the singlet operators. }:
$$
\eqalignno {
\Gamma^p_1(Q^2) & \equiv
\int_0^1 dx g_1^p(x;Q^2)  \cr
&= {1\over 6}  \biggl[ \biggl(G_A^{(3)}(0)
+ {1\over \sqrt3}G_A^{(8)}(0)\biggr)
\biggl(1-
{\a_s\over\p}-
3.583\Bigl({\a_s\over\p}\Bigr)^2
-20.215\Bigl({\a_s\over\p}\Bigr)^3 \biggr) \cr
&~~~~~
+ {2\over3} G_A^{(0)}(0;Q^2)
\biggl(1 - {1\over3}{\a_s\over\p}
-0.550\Bigl({\a_s\over\p}\Bigr)^2
\biggr)  \biggr]
, &(2.1) \cr }
$$
where the $G_A^{(a)}$ are form factors in the proton matrix elements
of the axial current
$$
\langle P|J^a_{\m5R}(k)|P\rangle ~=~
G_A^{(a)}(k^2) \bar u \c_\m \c_5 u ~+~
G_P^{(a)}(k^2) k_\m \bar u \c_5 u
,\eqno(2.2)
$$
and $a$ is an $SU(3)$ flavour index. In our normalisations (see ref.[7]):
$$
\eqalignno {
G^{(3)}_A &= {1\over2}(\Delta u-\Delta d) \cr
G^{(8)}_A &= {1\over{2\sqrt 3}}(\Delta u+\Delta d-2\Delta s) \cr
G^{(0)}_A &= \Delta u+\Delta d+\Delta s \equiv \Delta \Sigma
.&(2.3)  \cr }
$$
We ignore heavy quarks and, for simplicity, set the light
quark masses to zero in the formulae below.

The axial current occurs here since it is the lowest-twist,
lowest spin, odd-parity
operator in the OPE of two electromagnetic currents, i.e.
$$
J_\m(q)~J_\n(-q) ~~\bigQ~~2\sum_{a=0,3,8}
\e_{\m\n\a}{}^\b{q^\a \over Q^2}~C^a(Q^2)~J^a_{\b5R}
{}~+~ \ldots
.\eqno(2.4)
$$
The suffix $R$ emphasises that the current is the {\it renormalised}
composite operator. Under renormalisation, the gluon topological
density $Q_R$ and the divergence of the flavour singlet axial
current $J^0_{\m5R}$ mix as follows[12],
$$\eqalignno{
J^0_{\m5R} ~&=~ Z J^0_{\m5B}  \cr
Q_R ~&=~ Q_B ~-~ {1\over2N_F} (1-Z) \partial^\m J^0_{\m5B}
,&(2.5) \cr }
$$
where $J^0_{\m5B} = \sum \bar q \c_\m \c_5 q$ and
$Q_B = {\a_s\over8\p} \tr G^{\m\n}\tilde G_{\m\n}$
and we have quoted the formulae for $N_F$ flavours.
The mixing is such that the combination occurring in the axial anomaly
Ward identities, e.g.
$$
\langle0|\Bigl(\partial^\m J^0_{\m5R} - 2N_F Q_R \Bigr) ~ \tilde \OO_k
|0\rangle ~+~ \langle0|\d_5\tilde\OO_k |0\rangle ~~=~~0
,\eqno(2.6)
$$
is not renormalised.

Since $J^0_{\m5R}$ is renormalised, its matrix elements satisfy
renormalisation group equations with an anomalous dimension
$\c$, so that in particular $G_A^{(0)}(0;Q^2)$ depends on the RG scale
(which is set to $Q^2$ in eq.(2.1)).

As we have emphasised elsewhere, $G_A^{(0)}$ does {\it not},
as was initially supposed, measure the spin of the quark constituents
of the proton.The RG non-invariance of $J_{\m5R}^0$
(a consequence of the anomaly) is itself sufficient to prevent
this identification. The interest in the first EMC data [1,2]
on polarised $\m p$ scattering\footnote{$^*$}{\eightpoint
\noindent The combined SLAC/EMC data quoted in ref.[1] gives
$$
\C_1^p(Q^2=11\GV^2) ~=~ 0.126 \pm 0.010 \pm 0.015
$$
The result for $G_A^{(0)}$ in eq.(2.7) is extracted from the sum rule
using the values for $F$ and $D$ given above and the running coupling
from tau-decay data[30] (see the remarks after eq.(3.32)). },
which allows the following result for $G_A^{(0)}$ to be deduced,
$$
G_A^{(0)}(0;Q^2=11\GV^2) \equiv \Delta \Sigma~=~0.19 ~\pm~0.17
,\eqno(2.7)
$$
is rather that this value for $G_A^{(0)}$ represents a substantial
violation of the OZI rule[13,14], according to which we would expect
$$
G_A^{(0)}(0)_{OZI} = 3F-D \simeq 0.579 \pm 0.021
.\eqno (2.8)
$$
Here, we have used [15,16]:
$$
F+D\simeq 1.257 \pm 0.008 ~~~~~~~~~~
F/D \simeq 0.575 \pm 0.016
\eqno(2.9)
$$
as fitted from hyperon and $\beta$ decays.
The assumption that the OZI
rule is satisfied for $G_A^{(0)}(0)$ is equivalent to the
Ellis-Jaffe sum rule prediction for the first moment of $g_1^p$.

It follows immediately from eq.(2.2) (assuming the absence of a
massless pseudoscalar boson in the $U(1)$ channel) that
$$
G_A^{(0)}(0;Q^2)~\bar u \c_5 u ~=~
{1\over 2M}~\langle P|\partial^\m J^0_{\m5R} |P\rangle
,\eqno(2.10)
$$
where $M$ is the proton mass. The anomalous chiral Ward identity
then allows $G_A^{(0)}$ to be re-expressed as the forward
matrix element of the renormalised gluon topological density $Q_R$,
i.e.
$$
G_A^{(0)}(0;Q^2)~\bar u \c_5 u ~=~
{1\over 2M} 2N_F~\langle P| Q_R(0) |P\rangle
.\eqno(2.11)
$$
Notice that in terms of bare fields, $Q_R$ contains both gluon
and quark bilinears. This, together with the explicit factor
of $\a_s$ in the definition of the topological density, is the
source of the difficulty in giving a natural and unambiguous
parton interpretation[5,3,4].

At this point, we apply the method described in the
introduction.
We choose as the composite operator basis $\tilde\OO_k$ the
set of renormalised flavour singlet pseudoscalar operators,
viz.~$Q_R$ and and $\Phi_{5R}$, where,
up to a crucial normalisation factor discussed below,
the corresponding bare operator is simply the singlet
$i\sum\bar q \c_5 q$.
We then define $\C[Q_R, \Phi_{5R}; P, \bar P]$ to be the generating
functional of proper vertices which are 1PI with respect to these
composite fields only. (Here, $P$ and $\bar P$ denote interpolating
fields for the proton -- they play a purely passive r\^ole in the
construction.) $\C$ is obtained from the QCD generating functional
by a Legendre transform with respect to the sources for the
composite operators $Q_R$ and $\Phi_{5R}$ only. We may then write
(c.f. eq.(1.3))
$$
\langle P|Q_R(0)|P\rangle~~=~~
\langle 0|Q_R(0)~Q_R(0)|0\rangle~\C_{Q_R P \bar P} ~+~
\langle 0|Q_R(0)~\Phi_{5R}(0)|0\rangle~\C_{\Phi_{5R} P \bar P}
,\eqno(2.12)
$$
where the propagators are at zero momentum.

The composite operator propagator in the first term in eq.(2.12)
is the zero-momentum limit of an important quantity in QCD known
as the topological susceptibility $\chi(k^2)$, viz.
$$
\chi(k^2) ~=~\int dx~e^{ik.x} i\langle 0|T^* Q_R(x)~Q_R(0)|0\rangle
.\eqno(2.13)
$$
The second term is clearly independent of the normalisation of
the renormalised quark bilinear operator $\Phi_{5R}$.
We choose to normalise this operator in such a way that the inverse
two-point function $\C_{\Phi_{5R} \Phi_{5R}}$,
which has to vanish at $k^2 =0$, is equal to $k^2$, the correct
normalisation for a free, massless particle. With this normalisation,
a straightforward but intricate argument[7] using chiral Ward identities
(see Appendix A) shows that
the propagator $\langle 0|Q_R~\Phi_{5R}|0\rangle$ at zero momentum
is simply the square root of the first moment of the topological
susceptibility $\chi(k^2)$. We therefore find,
$$
\langle P|~Q_R(0)~|P\rangle ~~=~~\chi(0)~\C_{Q_R P \bar P}
{}~+~\sqrt{\chi^{\prime}(0)}~\C_{\Phi_{5R} P \bar P}
.\eqno(2.14)
$$

The chiral Ward identities further show that for QCD with massless
quarks, $\chi(0)$ actually vanishes. (This is in contrast to pure
Yang-Mills theory, where $\chi(0)$ is non-zero and is related to
the $\h^{\prime}$ mass in the large $N_C$ resolution of the $U(1)$
problem[17,18].)
Only the second term in eq.(2.14) remains.
Remarkably, this means that the matrix element of the renormalised
{\it gluon} density $Q_R$ measures the coupling of the proton to
the renormalised pseudoscalar {\it quark} operator $\Phi_{5R}$.
This happens because the composite operator propagator matrix
in the pseudoscalar ($Q_R, \Phi_{5R}$) sector is off-diagonal.
We therefore arrive at our basic result[7],
$$
G_A^{(0)}(0;Q^2)~\bar u \c_5 u
{}~~=~~ {1\over 2M} 2N_F~\sqrt{\chi^{\prime}(0)}~
\C_{\Phi_{5R} P \bar P}
.\eqno(2.15)
$$

The renormalisation group properties of eq.(2.15) are central to our
argument. With the normalisation of $\Phi_{5R}$ chosen above, it can be
shown[7] that the proper vertex $\C_{\Phi_{5R} P \bar P}$ is
RG invariant and so has no scale dependence.
The scale dependence needed to match $G_A^{(0)}$ is provided
entirely by the topological susceptibility which, as shown in
Appendix A, satisfies the RGE
$$
\biggl( \m {\partial\over\partial\m} +
\b(\a_s)\a_s {\partial\over\partial\a_s} - 2\c \biggr) \chi^{\prime}(0)
{}~=~0
.\eqno(2.16)
$$

The challenge posed by the EMC data is to understand the origin of the
OZI violation in $G_A^{(0)}$. The OZI approximation applied to
the r.h.s.~of eq.(2.15) would require\footnote{$^*$}{\eightpoint
\noindent To understand this, we note from ref.[7] that eq.(2.15)
is equivalently written as one form of the $U(1)$ Goldberger-Treiman
relation, viz.
$$
G_A^{(0)}(0;Q^2) ~=~ F_{\h_{OZI}} g_{\h_{OZI} NN}
$$
where $F_{\h_{OZI}}$ and $g_{\h_{OZI} NN}$ are respectively the
decay constant and nucleon coupling of a state $|\h_{OZI}\rangle$.
$|\h_{OZI}\rangle$ is an {\it unphysical} state in QCD (i.e.~not
a mass eigenstate) which in the OZI or large $N_C$ limit, in which
the anomaly is absent, can be identified as the massless $U(1)$
Goldstone boson. Simple quark counting rules then relate
$g_{\h_{OZI} NN}$ to the $\eta_8$-nucleon coupling $g_{\eta_8 NN}$.

This identification is the origin of our choice
of normalisation of $\Phi_{5R}$. In the OZI limit,
$\C_{\Phi_{5R} N \bar N}$ becomes the Goldstone
boson - nucleon coupling.}
(neglecting flavour $SU(3)$ breaking)
$\C_{\Phi_{5R} P \bar P} \simeq \sqrt{2} g_{\eta_8 NN} ~\bar u \c_5 u$
while $\sqrt{\chi^{\prime}(0)} \simeq (1/\sqrt{6}) f_\p$.

Our proposal is that we should expect the source of the OZI violation
to lie in RG non-invariant terms, i.e. in $\chi^{\prime}(0)$.
The reasoning is straightforward. In the absence of the $U(1)$
anomaly, the OZI rule would be an exact property of QCD. So the OZI
violation is a consequence of the anomaly. But it is the existence
of the anomaly that is responsible for the non-conservation and
hence non-trivial renormalisation of the axial current $J^0_{\m5R}$.
We therefore expect to find OZI violations in quantities sensitive
to the anomaly, which we identify through their RG dependence
on the anomalous dimension $\c$.
This seems reasonable since, if the OZI rule were to be good for
such quantities, it would mean approximating a RG non-invariant,
scale-dependent quantity by a scale-independent one.
%\vfill\eject
If this proposal is correct, we expect $\sqrt{\chi^{\prime}(0)}$ to be
significantly suppressed relative to its OZI approximation
of $(1/\sqrt{6}) f_\p$. The proper vertex $\C_{\Phi_{5R} P \bar P}$
would behave exactly as expected according to the OZI rule.
That is, the Ellis-Jaffe violating suppression of the first moment
of $g_1^p$ observed by EMC would {\it not} be a property of the
proton at all, but would simply be due to an anomalously small
value of the first moment of the QCD topological susceptibility
$\chi^{\prime}(0)$.

In the next section, we attempt to verify this hypothesis by
evaluating $\chi^{\prime}(0)$ using QCD spectral sum rules.

\vskip1cm

\noindent{\bf 3. QCD Spectral Sum Rule Estimate of $\chi^{\prime}(0)$}
\vskip0.5cm
We now present an estimate of
$\chi^{\prime}(0)$ in QCD with massless quarks
using the method of QCD spectral sum rules (QSSR) pioneered by
Shifman, Vainshtein and Zakharov
[19] and reviewed recently in ref.[20].

The correlation function $\chi(k^2)$ is defined in eq.(2.13)
and its renormalisation group equation is given in Appendix A.
Including the inhomogeneous contact term[21], we have
$$
\biggl(\m {\pl\over\pl \m} + \b(\a_s) \a_s
{\pl\over\pl\a_s} - 2\c \biggr)
\chi(k^2) ~=~ - {1\over(2N_F)^2} 2\b^{(L)} k^4
\eqno(3.1)
$$
with the beta function
$$
\b(\a_s)~\equiv~ {1\over \a_s} \m{d\over d\m} \a_s
{}~=~\b_1  \Bigl({\a_s\over\p}\Bigr)  ~+~
\b_2  \Bigl({\a_s\over\p}\Bigr)^2
,\eqno(3.2)
$$
where, for QCD with $N_F$ flavours,
$\b_1 = -{1\over2} \bigl(11-{2\over3}N_F\bigr)$
and $\b_2 = -{1\over4} \bigl(51 - {19\over3}N_F\bigr)$,
and the anomalous dimension[12]
$$
\c ~\equiv~ \m{d\over d\m} \log Z ~=~ -  \Bigl({\a_s\over\p}\Bigr)^2
.\eqno(3.3)
$$
The extra RG function $\b^{(L)}$ (so called because it appears in
the longitudinal part of the Green function of two axial currents)
is given by
$$
{1\over (2N_F)^2} \b^{(L)} ~=~ - {1\over 32\p^2} \Bigl(
{\a_s\over\p}\Bigr)^2 \biggl[1 + {29\over4} \Bigl({\a_s\over\p}\Bigr)
\biggr]
.\eqno(3.4)
$$
The RGE is solved in the standard way, giving
$$\eqalignno{
\chi(k^2,\a_s;\m)~~&=~~
e^{-2\int_0^t dt^{\prime}  \c\bigl(\a_s(t^{\prime})\bigr) } \biggl[
{}~\chi\bigl(k^2, \a_s(t);\m e^t\bigr)  \cr
&\hskip3.5cm
-~2\int_0^t dt^{\prime\prime} \b^{(L)}\bigl(\a_s(t^{\prime\prime})\bigr)
{}~e^{ 2\int_0^{t^{\prime\prime}} dt^{\prime}
\c\bigl(\a_s(t^{\prime})\bigr) } \biggr]
,&(3.5) \cr }
$$
where $\a_s(t)$ is the running coupling.

The perturbative expression for the two-point correlation function
in the $\overline{MS}$ scheme is[22]
$$
\chi(k^2)_{\rm P.T.}~\simeq~ -
\Bigl({\a_s\over8\p}\Bigr)^2 {2\over\p^2} k^4
 \log{-k^2\over\m^2}\biggl[1+\Bigl({\a_s\over \p}\Bigr)\biggl(
{1\over2}\b_1\log{-k^2\over\m^2} + {29\over4}\biggr) + \ldots \biggr]
.\eqno(3.6)
$$
The non-perturbative contribution from the gluon condensates
(coming from the next lowest dimension operators in the OPE)
is[23]
$$
\chi(k^2)_{\rm N.P.} ~\simeq~
- {\a_s\over 16\p^2} \biggl[\biggl(1 + {1\over2}
\b_1\Bigl({\a_s\over\p}\Bigr)
\log {-k^2\over\m^2}\biggr) \langle \a_s G^2\rangle
{}~-~2{\a_s\over k^2} \langle gG^3\rangle \biggr]
.\eqno(3.7)
$$
The RGE has been used to check the consistency of the leading log
approximation in the perturbative expression and to fix the radiative
correction in the gluon condensate contribution.

For the QSSR analysis of $\chi^{\prime}(0)$, we use the subtracted
dispersion relations
$$
{1\over k^2}\Bigl( \chi(k^2) - \chi(0) \Bigr)
{}~~=~~ \int_0^\infty {dt\over t}
{1\over t - k^2 - i\e} {1\over\p} {\rm Im}\chi(t)
\eqno(3.8)
$$
and
$$
{1\over k^4}\Bigl( \chi(k^2) - \chi(0) -k^2\chi^{\prime}(0)\Bigr)
{}~~=~~ \int_0^\infty {dt\over t^2}
{1\over t - k^2 - i\e} {1\over\p} {\rm Im}\chi(t)
.\eqno(3.9)
$$
Then, taking the inverse Laplace transform[20]
of both sides of the
dispersion relations and using the fact that $\chi(0) = 0$
in massless QCD, we find\footnote{$^*$}{\eightpoint
\noindent For the corresponding results in pure Yang-Mills theory,
see refs.[24,25]. }
$$\eqalignno{
&\int_0^{t_c}{dt\over t} e^{-t\t}{1\over\p} {\rm Im}\chi(t)  \cr
&\simeq~~\Bigl({\bar \a_s\over 8\p}\Bigr)^2 {2\over\p^2} \t^{-2}
\Bigl(1 - e^{-t_c\t}(1 + t_c\t)\Bigr) \biggl[1
{}~+ \Bigl({\bar \a_s\over 4\p}\Bigr)\biggl(29 + 4\b_1(1-\c_E)
- 8{\b_2\over\b_1}\log \bigl(-\log\t\L^2\bigr)\biggr)\biggr] \cr
&+ \Bigl({\bar\a_s\over8\p}\Bigr)\biggl[{1\over2\p}
\langle \a_s G^2\rangle + \Bigl({\bar\a_s\over\p}\Bigr) \t
\langle gG^3\rangle \biggr]
&(3.10) \cr }
$$
and
$$\eqalignno{
\chi^{\prime}(0)  ~~&\simeq ~~
\int_0^{t_c}{dt\over t^2} e^{-t\t} {1\over\p} {\rm Im}\chi(t)  \cr
%&\hskip0.5cm
&- \Bigl({\bar\a_s\over8\p}\Bigr)^2 {2\over\p^2} \t^{-1} \bigl(1 -
e^{-t_c\t}  \bigr)\biggl[1 + \Bigl({\bar\a_s\over4\p}\Bigr)
\biggl(29 - 4\b_1\c_E - 8{\b_2\over\b_1} \log\bigl(-\log \t\L^2
\bigr)\biggr)\biggr] \cr
&+\Bigl({\bar\a_s\over8\p}\Bigr)\biggl[{1\over2\p} \t
\langle \a_s G^2\rangle + \Bigl({\bar\a_s\over2\p}\Bigr)\t^2
\langle gG^3\rangle\biggr]
.&(3.11) \cr }
$$
where $\bar\a_s$ is the running coupling expressed in terms of
the QCD scale $\L$ from the two-loop relation:
$$
{\bar \a_s^{(2)}
\over\p} ~=~ \biggl({\bar\a_s\over\p}\biggr)\biggl[
1 - \biggl({\bar\a_s\over\p}\biggr) {\b_2\over\b_1}
\log \bigl(-\log \t \L^2 \bigr)  \biggr]
,\eqno(3.12)
$$
with
$$
{\bar\a_s\over \p}
{}~=~   {2\over \b_1 \log \t\L^2 }
.\eqno(3.13)
$$
In these expressions, we have cut off the $t$ integration at some
scale $t_c$ and used the perturbation theory approximation to
${\rm Im} \chi(t)$ for $t>t_c$.

In order to extract a value for $\chi^{\prime}(0)$ from these sum rules,
we keep only the lowest resonance (the $\eta^{\prime}$)
contribution to the spectral function, i.e. we assume
$$
{1\over\p} {\rm Im}\chi(t)~~=~~2 \tilde m^4_{\hp}
f_{\hp}^2 \d\bigl(t-\tilde m^2_{\hp}    \bigr)
{}~+~{\rm ``QCD~continuum"}~\theta(t-t_c)
,\eqno(3.14)
$$
where $\tilde m_{\hp} $ is the mass of the $\hp$ extrapolated
for massless QCD, viz.
$$
\tilde m^2_{\hp}    ~\simeq~m_{\hp}^2   - {2\over3}m_K{}^2
{}~\simeq~ (0.87 \GV)^2
.\eqno(3.15)
$$
To evaluate eqs.(3.10) and (3.11),
we use
$$
\L~\simeq~(350 \pm 100)\MV
\eqno(3.16)
$$
for the QCD scale parameter[26],
$$
\langle \a_s G^2\rangle ~\simeq~ (0.06 \pm 0.02)\GV^4
\eqno(3.17)
$$
from a global fit of the light mesons and charmonium data[20],
and parametrise the triple gluon condensate as
$$
\langle g^3 G^3\rangle ~\simeq~(1.5 \pm 0.5)\GV^2
\langle \a_s G^2\rangle
\eqno(3.18)
$$
using the dilute gas instanton model[19].
We show the result in Fig.5a for $\chi^{\prime}(0)$ plotted
versus $\t$ for different values of $t_c$.
In Fig.5b we show the behaviour of the $\t$ minima for different
$t_c$. Our optimal result corresponds to the range of values of $t_c$
corresponding to the first appearance of the $\t$ minimum until
the beginning of the $t_c$ stability region.
The value of $\t$ at which the stability occurs is around
$0.4$ to $0.6 \GV^2$, which is quite small compared with the light
meson systems and is consistent with qualitative expectations[23]
of a scale hierarchy in the QSSR analysis of gluonium systems.
This small value of $\t$ also ensures that higher-dimension
operators such as those arising from instanton-like effects
will not contribute in the OPE. We deduce:
$$
\sqrt{\chi^{\prime}(0)}~~\simeq~~(22.3\pm3.2\pm2.8\pm1.3)\MV
,\eqno(3.19)
$$
where the first error comes from $\langle \a_s G^2\rangle$, the
second one from $\L$ and the third from the range of $t_c$ values
from $4.5\GV^2$ to $7.5\GV^2$. The effects of the triple gluon
condensate and the radiative corrections are relatively unimportant,
contributing about $3-10\%$ to $\chi^{\prime}(0)$.
We add a guessed error of $5\%$ each from the unknown non-perturbative
and radiative correction terms. Finally, adding all these errors
quadratically, we find the following Laplace sum rule
estimate of the first moment of the topological susceptibility
evaluated at $\t \simeq 0.5\GV^{-2}$:
$$
\sqrt{\chi^{\prime}(0)}~~\simeq~~(22.3 \pm 4.8) \MV
.\eqno(3.20)
$$

\vskip0.3cm
As a check on the validity of this result, we now repeat the analysis
using the finite enrgy sum rule
(FESR) local duality version of the spectral sum rules
discussed in ref.[25].
The advantage of the FESR method is that it projects out the effects
of the operators of a given dimension[27] (in this case, dimension 4)
in such a way that, at the order to which we are working,
the FESR analogues of the sum rules (3.10) and (3.11) are not
affected by higher dimension operators such as those induced by
instanton-like effects.

The FESR sum rules are
$$\eqalignno{
\int_0^{t_c} {dt\over t} {1\over\p} {\rm Im}\chi(t) ~~&\simeq~~
\Bigl({\bar\a_s\over8\p}\Bigr)^2 {2\over\p^2} {t_c^2\over2}
\biggl[1 + \Bigl({\bar\a_s\over4\p}\Bigr)
\biggl(29 - 4{\b_1\over2} - 8
{\b_2\over\b_1} \log\bigl(-\log \t\L^2
\bigr)\biggr)\biggr] \cr
&  +~~ \Bigl({\bar\a_s\over8\p}\Bigr)\biggl[{1\over2\p}
\langle \a_s G^2\rangle \biggr]
&(3.21) \cr }
$$
and
$$
\eqalignno{
\chi^{\prime}(0) ~~ &\simeq~~
\int_0^{t_c} {dt\over t^2} {1\over\p} {\rm Im}\chi(t) \cr
{}~&-~~ \Bigl({\bar\a_s\over8\p}\Bigr)^2 {2\over\p^2} t_c
\biggl[1 + \Bigl({\bar\a_s\over4\p}\Bigr)\biggl(
29 - 4\b_1 - 8{\b_2\over\b_1}\log \bigl(-\log \t\L^2\bigr)\biggr)
\biggr]
&(3.22) \cr}
$$
Analysing eqs.(3.21) and (3.22), we realise that the solution
increases monotonically with $t_c$ so that no firm prediction
can be made, although the result gives a rough indication of
consistency with the previous Laplace sum rule.
To overcome this problem, we repeat the analysis using only the
FESR (3.22) and using as an extra input the value of the parameter
$f_{\hp}$ extracted from the first Laplace sum rule (3.10).
The value of $f_{\hp}$ is given in Appendix B.
This weakens the $t_c$ dependence of the result and $t_c$ stability
now appears as an inflection point.
We obtain the result shown in Fig.6 for different values of
$f_{\hp}$ and $\L$, from which we deduce that with $t_c \simeq
6.5-9.5\GV^2$,
$$
\sqrt{\chi^{\prime}(0)}~\simeq~ (25.5\pm1.5\pm2.0\pm1.0 )\MV
,\eqno(3.23)
$$
where the errors come from $f_{\hp}$, $\L$ and $t_c$ respectively.
Adding the errors quadratically and including a further $5\%$
error from the unknown higher order terms, we obtain
at the scale $\t \simeq 0.5 \GV^{-2}$:
$$
\sqrt{\chi^{\prime}(0)}~\simeq~(26.5 \pm 3.1)\MV
,\eqno(3.24)
$$
where, we have run the result from $t_c = 8 \GV^2$
to the scale $\t = 0.5\GV^{-2}$ using the
RGE solution expressed in terms of $\L$, viz.
$$
\chi^{\prime}(0;\m) ~\simeq~ \hat \chi^{\prime}(0)
e^{8\over \b_1^2 \log \m/\L}
,\eqno(3.25)
$$
where $\hat \chi^{\prime}(0)$ is RG invariant.
(Notice that the inhomogeneous term proportional to $\b^{(L)}$
does not contribute to the first moment at $k^2=0$.)
We see that the FESR result is consistent with the Laplace one.

\vskip0.3cm
Taking the average of the Laplace and FESR results, we obtain
our final estimate of the first moment of the topological
susceptibility at the scale $\t = 0.5\GV^{-2}$:
$$
\sqrt{\chi^{\prime}(0)} ~\simeq~(25.3\pm2.6)\MV
.\eqno(3.26)
$$
This result should be compared with that obtained[24,25] in pure $N_C=3$
Yang-Mills theory using a similar QSSR approach:
$$
\sqrt{-\chi^{\prime}(0)}\big|_{\rm YM} ~\simeq~(7 \pm 3)\MV
.\eqno(3.27)
$$
It is important to notice that this pure Yang-Mills result has
been confirmed by lattice calculations[28,29], which is a strong
indication of the validity of the methods used in deriving both
(3.27) and (3.26). The introduction of massless quarks
has changed the sign of $\chi^{\prime}(0)$ and increased its
absolute value by a factor of around 4. From the QSSR analysis, this
effect is due mainly to the low value of the $\hp$ mass of
$0.87\GV$ (massless QCD) which enters into the spectral function,
compared with the pseudoscalar gluonium
mass of about $1.36-1.66\GV$ [24,20] in pure Yang-Mills theory.

To compare with the experimental result on the polarised proton
structure function, we use the RGE to run the result for
$\chi^{\prime}(0)$ to the EMC scale of $10\GV^2$. We find
$$
\sqrt{\chi^{\prime}(0)}\big|_{EMC} ~\simeq~
(23.2\pm2.4)\MV
.\eqno(3.28)
$$
This is smaller by a factor of $1.64\pm0.17$ than the OZI value
of $(1/\sqrt6)f_\p$.
We therefore do indeed find a significant
suppression of $\chi^{\prime}(0)$
relative to its OZI value.

To convert this result into a prediction for the singlet form factor,
we take our fundamental expression (2.15) for $G_A^{(0)}$
and equate the proper vertex $\C_{\Phi_{5R} P \bar P}$ with
its OZI expression given by the Goldstone boson-nucleon coupling. In
this way, we obtain:
$$
G^{(0)}_A(0)~=~G^{(0)}_A(0)_{OZI} ~ {\sqrt{\chi^{\prime}(0)}
\over {(1/\sqrt6)f_\p} }
\eqno(3.29)
$$
Using the value  of $G^{(0)}_A(0)_{OZI}$ in eq.(2.8)
and including an additional error of approx. $10\%$ for the
use of the OZI approximation for the proper vertex, we arrive at our
final prediction:
$$
G_A^{(0)}(0;Q^2=10\GV^2) ~\simeq~ 0.353 \pm 0.052
.\eqno(3.30)
$$
Substituting this result\footnote{$^*$}{\eightpoint
\noindent In terms of the quantities $\D u$, $\D d$ and $\D s$ defined
in eq.(2.3), we have at $Q^2=10\GV^2$:
$$
\D u ~=~ 0.84 \pm 0.01~~~~~~
\D d ~=~ - 0.41 \pm 0.01~~~~~~
\D s ~=~ - 0.08 \pm 0.02
$$ }
together with
$$ \eqalignno {
G_A^{(8)} &\equiv {1\over 2\sqrt {3}}(3F-D) \cr
G_A^{(3)}  &\equiv {1\over 2}(F+D)
,&(3.31) \cr }
$$
into the first moment sum rule (2.1), using the values of $F$ and $D$
from eq.(2.9), and neglecting the higher twist terms (which are
certainly negligible at $Q^2=10~\GV^2$) we deduce:
$$
\Gamma_1^p(10 \GV^2) \simeq  0.143 \pm 0.005
.\eqno (3.32)
$$
Here, we have used the coupling $\alpha_s(m_{\tau})= 0.347 \pm 0.030$
extracted from tau-decay data[30]. One should
notice that the radiative corrections decrease the leading order result
by about 12$\%$.

Our result, eqs.(3.30) and (3.32), certainly goes in the right direction,
i.e.~that of reducing the prediction from the OZI (Ellis-Jaffe) value.
At the time we obtained it, however, eq.(3.30) still appeared too high
compared to the experimental result (2.7), which would have implied
further OZI violations in the proper vertex.
Amusingly enough, while this paper was being completed we learned
of the new results from the SMC collaboration which,
combined with the earlier proton data, gives the new world average [31]:
$$
\Gamma_1^p(10~\GV^2)= 0.145 \pm 0.008 \pm 0.011
,\eqno (3.33)
$$
from which we deduce
$$
G^{(0)}_A (0;Q^2=10\GV^2) \equiv \Delta \Sigma = 0.37 \pm 0.07 \pm 0.10
.\eqno (3.34)
$$
These results are now in excellent agreement with our predictions.

\vskip1cm

\noindent{\bf 4. Tests of the Bjorken sum rule and estimate of
higher twist effects}
\vskip0.5cm
Recently, the SMC collaboration at CERN[31],[32]
and the E142 collaboration at SLAC[33] have produced data on the
polarised neutron structure function $g_1^n$. Since our proposal
requires that the flavour singlet suppression is identical for
the proton and neutron, we see no reason why the Bjorken sum rule
[34]:
$$
\eqalignno {
\delta \Gamma^{p-n}_1  \equiv \Gamma^p_1-\Gamma^n_1 &\equiv
\int_0^1 dx \Bigl( g_1^p(x;Q^2) - g_1^n(x;Q^2) \Bigr) \cr
&~=~
{1\over 6} g_A \biggl(1 - {\a_s\over \p}
-3.583\Bigl({\a_s\over\p}\Bigr)^2
-20.215\Bigl({\a_s\over\p}\Bigr)^3 \biggr)
+ {{a_p-a_n}\over Q^2}
 &(4.1) }
$$
should not hold, at least up to
flavour $SU(2)$ breaking. Provided the measurements are at
sufficiently high $Q^2$, the higher twist corrections related to the
coefficients $a_p-a_n$ can be neglected.
Analysis of the combined proton and deuteron data as performed
in ref.[35] gives at $Q^2=5 \GV^2$ [31]:
$$
\delta\Gamma^{p-n}_1 \simeq 0.203 \pm 0.029
,\eqno (4.2)
$$
to be compared with the QCD prediction,
with $\a_s(5\GV^2) =  0.32 \pm 0.02 $, of
$$
\delta\Gamma^{p-n}_1\simeq (0.176 \pm 0.003)+
 {{a_p-a_n}\over Q^2}
,\eqno (4.3)
$$
{}From this, one can deduce the difference of the higher twist
coefficients (in units of $\GV^2$):
$$
a_p-a_n \simeq   0.135 \pm 0.145
. \eqno (4.4)
$$

We can pursue an analogous analysis for the first moment of the
neutron structure function, which satisfies the sum rule (c.f.~eq.(2.1)):
$$
\eqalignno {
\Gamma^n_1(Q^2) & \equiv
\int_0^1 dx g_1^n(x;Q^2)  \cr
&= {1\over 6}  \biggl[ \biggl(-G_A^{(3)}(0)
+ {1\over \sqrt3}G_A^{(8)}(0)\biggr)
\biggl(1- {\a_s\over\p}-
3.583\Bigl({\a_s\over\p}\Bigr)^2
-20.215\Bigl({\a_s\over\p}\Bigr)^3 \biggr) \cr
&~~~~~+ {2\over3} G_A^{(0)}(0;Q^2)
\biggl(1 - {1\over3}{\a_s\over\p}
-0.550\Bigl({\a_s\over\p}\Bigr)^2
\biggr)  \biggr] +{a_n \over Q^2}
, &(4.5) \cr }
$$
where we have included the higher twist contribution. Evaluating
this quantity at $Q^2=2\GV^2$, where the SLAC data are available,
we find
$$
\Gamma_1^n (2 \GV^2) \simeq -(0.031 \pm 0.006) +{a_n\over Q^2}
.\eqno(4.6)
$$
Comparing this with the SLAC data[33],
$$
\Gamma_1^n (2 \GV^2) \simeq -(0.022 \pm 0.011),
\eqno(4.7)
$$
and using eq.(4.4),
we can extract the coefficients of the higher twist terms.
In units of $\GV^2$, we find:
$$
\eqalignno {
a_p &\simeq  -0.117 \pm 0.145 \cr
a_n & \simeq ~0.018 \pm 0.025
.&(4.8) \cr}
$$

These values of the higher twist terms are consistent with the
previous determinations[36,37] from QCD spectral sum rules.
However, these sum rules would be affected by
a more general choice of the nucleon interpolating field[20] (the one
used in refs.[36,37] is not the optimal one) and
by the well-known[20] large violation
by a factor 2-3 of the vacuum saturation of the four-quark condensate,
which is assumed in refs.[36,37] to be satisfied to within 10-20$\%$.
In addition, radiative corrections, which are known to be large in
the baryon sum rules[20], can also be important here.
More accurate data on the Bjorken and neutron sum rules,
and/or a measurement of the proton sum rule at lower $Q^2$,
are needed to improve the results in eq.(4.8), which are necessary
to test the validity of the QCD spectral sum rule
predictions in ref.[37].

\vfill\eject

\noindent{\bf 5. Further Discussion}
\vskip0.2cm
In this paper, we have presented evidence that the experimentally
observed suppression of the first moment of the polarised proton
structure function $g_1^p$ (the so-called EMC ``proton spin'' crisis)
is a target-independent effect reflecting a suppression of the first
moment of the QCD topological susceptibility $\chi^{\prime}(0)$
relative to the OZI expectation. Not only does $G_A^{(0)}(0)$
not measure the quark spin -- its suppression is not even a
property of the proton structure.

It would be interesting to test this hypothesis directly by polarised
deep inelastic scattering experiments
on other targets not simply related to the proton
by flavour symmetry. We have already studied the case of
a photon target and have presented elsewhere[38] a new sum rule
for the first moment of the polarised photon structure function
$g_1^\c$ measurable in polarised $\el^+ \el^-$ colliders.
However, this turns out to be a special case because the
electromagnetic $U(1)$ anomaly contributes at leading order and
so the $g_1^\c$ sum rule does not display the suppression
mechanism described here. Another possibility is to consider
semi-inclusive processes in which a particular hadron with a
fraction $z$ of the incoming momentum is observed in the target
fragmentation region. It was recently suggested[39]
that such cross-sections should be described in terms of new,
non-perturbative hybrid functions $M(z,x,Q)$, called
``fracture functions''.
To the extent that an OPE can be used, it would be
possible to represent $M$ in terms of the forward matrix element of
a composite operator between a suitable proton-plus-hadron state. In this
case, one would again factorise $M$ into a composite propagator of the
usual type and a proper vertex involving four external hadron legs. If
the suppression of the polarised structure function indeed originates
from the propagator, as we suggest, such a suppression should also be
found at the level of the (less inclusive) fracture functions.

So far, we have only considered the first moment of $g_1^p$.
Of course, we would like to extend our approach to higher moments
and discuss the full $x$-dependence of the structure function.
This would require knowledge of the renormalisation properties
and composite operator Green functions of
the higher-spin axial currents and gluon densities[40], together with
the associated proper vertices.

Another possible line of development would be to try to develop
techniques to estimate the proper vertices themselves, rather than
just the composite operator Green functions.
To the extent that the quenched approximation may be trusted for the
proper vertices, lattice calculations could already be suitable for
the task, and QCD spectral sum rule techniques could be used in
conjunction to check the validity of that approximation. We
recall that, in contrast, use of the quenched approximation directly
for the matrix elements of the operator $Q$ can be shown to be completely
unreliable since these are affected by low-lying poles that should
disappear after dynamical quark loops are added. This is another example
of how the apparent complication introduced by our splitting of
matrix elements into propagators and proper vertices can ultimately
pay dividends.

Finally, it would be interesting to attempt to apply this analysis
of deep inelastic scattering using proper vertices to other
QCD processes normally described in the language of the parton
model rather than in terms of the OPE. Semi-inclusive deep inelastic
scattering is one such example,
but many other interesting possibilities can be considered, especially in
the context of hadron-hadron collisions.

\vfill\eject

{\vsize20cm
\noindent{\bf Appendix A~~~~Chiral Ward Identities and the
Renormalisation Group}
\vskip0.5cm
The anomalous chiral Ward identities for Green functions of the
pseudoscalar operators $Q_R$ and $\Phi_{5R}$ are (for zero
quark masses)
$$\eqalignno{
&ik_\m~\langle0|J_{\m5R}^0(k)~Q_R(-k)|0\rangle
{}~-~ 2N_F~\langle0|Q_R(k)~Q_R(-k)|0\rangle ~=~0
&(A.1) \cr
&ik_\m~\langle0|J_{\m5R}^0(k)~\Phi_{5R}(-k)|0\rangle
{}~-~ 2N_F~\langle0|Q_R(k)~\Phi_{5R}(-k)|0\rangle ~+~
\langle 0|\d_5 \Phi_{5R}(-k)|0\rangle ~=~0.
&{}    \cr
&{}&(A.2) \cr }
$$
So, at zero momentum, assuming there is no physical
massless $U(1)$ boson,
$$
\langle0|Q_R(0)~Q_R(0)|0\rangle ~=~0
,\eqno(A.3)
$$
showing that the topological susceptibility $\chi(0)$
vanishes for massless QCD, and
$$
\langle0|Q_R(0)~\Phi_{5R}(0)|0\rangle ~=~ -{1\over2N_F} 2\langle
\Phi_R\rangle
,\eqno(A.4)
$$
where $\langle \Phi_R \rangle$ is the VEV of the scalar partner
of $\Phi_{5R}$ and is non-vanishing because of the quark condensate.

The field $\Phi_{5R}$ is normalised such that the two-point
proper vertex $\C_{\Phi_{5R}\Phi_{5R}} = k^2$. This means that
$\C_{\Phi_{5R}\Phi_{5R}}$ is (minus) a component of the inverse
propagator matrix in the pseudoscalar sector, i.e.
$$
\C_{\Phi_{5R}\Phi_{5R}} ~=~
\langle0|Q_R~Q_R|0\rangle \biggl(
\langle0|Q_R~\Phi_{5R}|0\rangle^2 ~-~
\langle0|Q_R~Q_R|0\rangle \langle0|\Phi_{5R}~\Phi_{5R}|0\rangle
\biggr)^{-1}
.\eqno(A.5)
$$
Expanding to lowest order in $k^2$ gives
$$
\C_{\Phi_{5R}\Phi_{5R}} ~=~
\chi^{\prime}(0)~\langle0|Q_R(0)~\Phi_{5R}(0)|0\rangle^{-2}~k^2
{}~+~O(k^4)
,\eqno(A.6)
$$
where we have written $\langle0|Q_R(k)~Q_R(-k)|0\rangle
= \chi^{\prime}(0)k^2 + O(k^4)$. We therefore deduce
$$
\langle0|Q_R(0)~\Phi_{5R}(0)|0\rangle ~=~
\sqrt{\chi^{\prime}(0)}
,\eqno(A.7)
$$
as quoted in eq.(2.15).

\vskip0.3cm
The renormalisation group equation for the topological susceptibility
follows from the definition of the renormalised composite operators,
eq.(2.4), and the chiral Ward identities.
The Ward identity for the two-current Green function is
$$
ik^\m~\langle0|J_{\m5R}^0(k)~J_{\n5R}^0(-k)|0\rangle ~-~
2N_F~\langle0|Q_R(k)~J_{\n5R}^0(-k)|0\rangle ~=~ 0
.\eqno(A.8)
$$
Combining eqs.(A.1), (A.8) and (2.4), we find straightforwardly
$$
\langle0|Q_R(k)~Q_R(-k)|0\rangle ~=~
Z^2~\langle0|Q_B(k)~Q_B(-k)|0\rangle ~+~ \ldots
.\eqno(A.9)
$$
The dots denote the extra divergences associated with contact terms
in the two-point Green functions of composite operators. Taking these
into account (see refs.[21,7] for full details) we find the full
RGE for $\chi(k^2)$,
$$
\biggl(\m{\partial\over\partial\m} + \b(\a_s) \a_s
{\partial\over\partial \a_s} - 2\c \biggr) \chi(k^2) ~=~
-{1\over(2N_F)^2}2\b^{(L)}(\a_s) k^4
,\eqno(A.10)
$$
where $\b^{(L)}$ is a new RG function. The inhomogeneous term does not
contribute at zero momentum, however, and the required RGE (2.13)
for $\chi^{\prime}(0)$ follows immediately.
}
\vfill\eject

\noindent{\bf Appendix B ~~~~~~Decay Constants and the $\hp$ }
\vskip0.5cm
We can estimate the parameter $f_{\hp}$ appearing in the spectral
expansion using the first Laplace QSSR, eq.(3.10). $f_{\hp}$ is defined
by
$$
\langle 0|J_{\m5R}^0(k) |\hp\rangle ~=~ik_\m f_{\hp}
,\eqno(B.1)
$$
and is RG non-invariant.
On shell (see ref.[7], Appendix D), the scale dependence is due
entirely to the anomalous dimension $\c$ of the axial current so, using
eq.(3.3) and expressing the result in terms of the QCD scale $\L$,
we may write
$$
f_{\hp}(\m) ~=~ \hat f_{\hp} e^{4\over \b_1^2\log \m/\L}
,\eqno(B.2)
$$
where $\hat f_{\hp}$ is RG invariant.
{}From the QSSR (3.10), we find the $\t$ stability starts at $t_c
\simeq 6.5 \GV^2$, while the $t_c$ stability is reached for $t_c$
larger than 9.5$\GV^2$. In this region, the radiative corrections are
about 10$\%$ of the lowest order term, while the $<g^3 G^3>$ one
contributes about 10$\%$. Under such conditions, our optimal result
at $\t   \simeq 0.6 \GV^{-2} $ is (see Fig. 6a):
$$
f_{\hp} \simeq (24.1 \pm 0.6 \pm 3.4 \pm 0.3) {\rm  MeV}
,\eqno(B.3)
$$
where the first error comes from $<\a_s G^2>$, the second
from $\L$ and the third from the range of $t_c$ values
between 6.5 $\GV^2$ and 9.5 $\GV^2$. Adding a $5\%$ error from the
unknown QCD terms, adding the different errors quadratically
and running to the EMC scale, we obtain
$$
f_{\hp}|_{EMC} \simeq (23.6 \pm 3.5) {\rm  MeV}.
\eqno(B.4)
$$
This value is strongly suppressed relative to
the OZI prediction of $\sqrt 6 f_\p$ for the $\hp$ decay constant.

\vskip0.3cm
However, as has been shown in refs.[41,7], this $f_{\hp}$ is
{\it not} the $\hp$ decay constant measured in, e.g., the decay
$\hp \ra \c\c$. In fact, the analogues of the current algebra
formulae
$$
f_\p g_{\p\c\c}  ~=~ {  1\over \p} \a_{\rm em}
,\eqno(B.5)
$$
and
$$
f_\p g_{\p NN} ~=~ m_N g_A
,\eqno(B.6)
$$
in the flavour singlet sector are[41,7]
$$
F g_{\hp \c\c} + {1\over 2N_F} F^2 m_{\hp}^2 g_{G\c\c}(0) ~=~
{4   \over  \p} \a_{\rm em}
,\eqno(B.7)
$$
and
$$
F g_{\hp NN} + {1\over2N_F} F^2 m_{\hp}^2 g_{GNN}(0) ~=~
2m_N G_A^{(0)}(0)
.\eqno(B.8)
$$
Here, $F$ is the RG invariant decay constant defined by
$$
F~=~ {2 \langle\phi_R\rangle\over m_{\hp}}\biggl( \int dx~i \langle 0|T^*
\phi_{5R}^0(x)~\phi_{5R}^0(0) |0\rangle \biggr)^{-{1\over2}}
,\eqno(B.9)
$$
where $\phi_5^0 = i\sum \bar q \c_5 q$ and $\vev =
\sum \langle \bar q q \rangle$.
The extra terms $g_{G\c\c}$ and $g_{GNN}$ appearing in eqs.(B.7,B.8)
(which are properly defined as proper vertices[41,7]) may be
thought of as the couplings of the gluonic component of the $\hp$.
They arise because the $\hp$ is not a Goldstone boson in the $U(1)$
channel and so the naive current algebra extensions of eqs.(B.5,B.6)
are not valid. At first sight, therefore, eqs.(B.7) and (B.8)
are not predictive since $g_{G\c\c}$ and $g_{GNN}$ are unknown.
However, if we follow our proposal that OZI violations are
associated with RG non-invariant quantities we can make predictions.

Taking eq.(B.7) first, we have shown[41] that
$g_{G\c\c}$ is RG invariant.
Since in the OZI limit this term is absent, we therefore expect
$g_{G\c\c}$ to be small, and so to a good approximation we predict
$$
F g_{\hp \c\c} ~=~ {4   \over \p} \a_{\rm em}
.\eqno(B.10)
$$
Since $F$ is RG invariant, we expect it to be well approximated
by its OZI value $\sqrt 6 f_\p$. Experimentally (see ref.[42]),
the relation (B.10) is very well satisfied.

In eq.(B.8), on the other hand, $g_{GNN}$ is not RG invariant so we
do not expect this term to be small. In fact, this equation is just a
rewriting of the $U(1)$ GT formula quoted in the text, for which
our proposal is successful.

An important test of our picture of the pattern of OZI breaking is
therefore to evaluate the RG invariant decay constant $F$ from first
principles and check that it is close to the OZI prediction of
$\sqrt{2N_F} f_\p$. Again, we can use QCD spectral sum rules.
\vskip0.3cm

We require the zero-momentum limit $\Phi_5(0)$
of the two-point correlation function
$$
\Phi_5(k^2) ~=~ \int dx e^{ik.x} i\langle0|T^* \phi_{5R}^0(x)~
\phi_{5R}^0(0)|0\rangle
\eqno(B.11)
$$
for QCD with 3 flavours and massless quarks.
However, as there is a smooth behaviour of the two-point correlator
when the common light quark mass $m_R$ goes
to zero, we shall work (for convenience) with the RG invariant
correlation function
$$
\Psi_5(k^2) ~\equiv~ 4m_R^2 \Phi_5(k^2)
\eqno(B.12)
$$
where $m_R$ is the average of the renormalised u and d quark masses.
Now, in perturbation theory, the difference between this
flavour singlet correlation function and the corresponding
non-singlet one appears only at $O(\a_s^2)$ from the double-triangle
anomaly-type diagrams. Similarly for the non-perturbative
condensate terms, the difference is only of $O(\a_s^2)$ arising
from the equivalent diagrams. Instanton-like effects appear as
higher-dimension operators. So, at the order we are working,
we can simply use the expression for the isotriplet (pion)
correlation function in QCD discussed in the literature[20].

The first Laplace sum rule to two-loops reads[20]:
$$\eqalignno{
\int_0^{t_c} dt e^{-t\t} {1\over\p} {\rm Im} \Psi_5(t)
&~~\simeq~~
{3N_F\over2\p^2}\bar m^2(\t)
 \Biggl\{ \t^{-2}
\Bigl(1 - e^{-t_c \t}(1+t_c\t)\Bigr) \cr
&\biggl[1~-~ {2\over \b_1 L}
\biggl({11\over3} + 2\c_E
- {2\over \b_1}\biggl(\tc_2 - \tc_1{\b_2\over\b_1}\biggr)
+ 2 {\tc_1\b_2\over \b_1^2} \log L \biggr)\biggr] \cr
&+~~\biggl[{\p\over3} \langle \a_s G^2\rangle
+ {896\over81} \p^3 \r \a_s \langle \bar u u \rangle^2\t \biggr]
\Biggr\}
&(B.13) \cr }
$$
where $L= -\log \t \L^2$ and [20]:
$$
\eqalignno{
&\r \a_s \langle \bar u u \rangle^2
{}~\simeq~ (3.8 \pm 2.0) 10^{-4} \GV^6 \cr
& \bar m (\t) ~\equiv~ {1\over 2}
(\bar m_u+\bar m_d)(\t)~\simeq~
(-{1\over2}\log \t \L^2)^{ \tc_1/\b_1}
{}~(12.1 \pm 1.0) \MV
&(B.14) \cr}
$$

As before, we parametrise the spectral function keeping only the
lowest ($\hp$) resonance, i.e.
$$
{1\over\p} {\rm Im} \Psi_5(t) ~~=~~ 2 \tilde m_{\hp}^4 f^2
\d\bigl(t - \tilde m_{\hp}^2\bigr) ~+~
{\rm ``QCD~continuum "}~\theta(t-t_c)
,\eqno(B.15)
$$
where the unknown parameter $f$, which is defined by
$$
2m_R\langle0|\phi_{5R}^0|\hp\rangle ~=~
\sqrt{2}    f \tilde m^2_{\hp}
,\eqno(B.16)
$$
can be estimated from the sum rule (B.13).
We study the $\t$ and $t_c$ behaviours of $f$ in Fig.7. The $\t$
stability starts for $t_c \simeq 4\GV^2$, while stability in $t_c$
appears above $t_c \simeq 7\GV^2$, a range which is equal to the one
for the correlation function for $Q(x)$. The value for the $\t$
stability of about
$0.9 \GV^{-2}$
is typical of light quark correlation functions.
At the minimum, we obtain
$$
f~=~ \sqrt{N_F}~   \bigl(5.55 \pm 0.08 \pm 0.65 \pm
0.35 \pm 0.06 \pm 0.03  \bigr) \MV
,\eqno(B.17)
$$
where the errors come respectively from $t_c$,
$\L$,
$\bar m$,
$\langle \a_s G^2 \rangle$ and $\r \a_s \langle \bar u u\rangle^2$.
Adding these errors quadratically, we deduce
$$
f ~=~   \sqrt{N_F}~ (5.55\pm 0.75) \MV
.\eqno(B.18)
$$

With this value for $f$, we are now able to estimate $\Psi_5(0)$
itself using a second Laplace sum rule[43,25]:
$$\eqalignno{
\Psi_5(0) ~~&\simeq~~\int_0^{t_c} {dt\over t}
e^{-t\t} {1\over\p} {\rm Im} \Psi_5(t)
-~~{3N_F
\over2\p^2}
\bar {m}^2(\t)
\Biggl\{
\t^{-1} \Bigl(1 - e^{-t_c \t}\Bigr) \cr
&\biggl[1 ~-~  {2\over \b_1 L}\biggl({11\over3} + 2\c_E
- {2\over \b_1}\biggl(\tc_2 - \tc_1{\b_2\over\b_1}\biggr)
+ 2 {\tc_1\b_2\over \b_1^2} \log L \biggr)\biggr] \cr
&+~~ \biggl[{\p\over3}\langle \a_s G^2\rangle
+ {1\over2}~ {896\over 81} \p^3 \r \a_s \langle \bar u u\rangle^2
\t \biggr] \Biggr\}
&(B.19) \cr}
$$
where $\tc_1$, $\tc_2$ are the coefficients in the anomalous
dimension for the light quark mass. For three flavours,
$\tc_1= 2$ and $\tc_2=91/12$.
The sum rule analysis of this quantity shows a strong $t_c$
dependence and the $\t$ stability only appears at unrealistic
values of $t_c$ larger than $8\GV^2$.
In order to circumvent this difficulty, we work with a combination
of the sum rules (B.13) and (B.19) which has been used
successfully in the past for measuring the deviation from
pion and kaon PCAC to a good accuracy[43].
The combined sum rule reads:
$$\eqalignno{
\Psi_5(0) ~~&\simeq~~\int_0^{t_c} {dt\over t} e^{-t\t}(1-t\t)
{1\over\p}{\rm Im}\Psi_5(t)
-~~{3N_F \over2\p^2}\bar m^2(\t)
\Biggl\{ \t^{-1}
\Bigl(t_c\t~e^{-t_c \t}\Bigr) \cr
&\biggl[1 ~-~  {2\over \b_1 L}
\biggl({11\over3} + 2\c_E
- {2\over \b_1}\biggl(\tc_2 - \tc_1{\b_2\over\b_1}\biggr)
+ 2 {\tc_1\b_2\over \b_1^2} \log L \biggr)\biggr] \cr
&+~~ \t\biggl[{2\p\over3}\langle\a_sG^2\rangle
+ {3\over2}~{896\over81} \p^3\r\a_s\langle\bar u u\rangle^2 \t
\biggr] \Biggr\}
&(B.20) \cr}
$$
This sum rule is studied in Fig.8. The position of the
stability is almost insensitive to the value of $t_c$ due
to some cancellations amongst the perturbative terms.
However, this feature also implies that the stability is
obtained at values of $\t$ larger than in the previous cases
making the result sensitive to the errors on the four-quark
condensates, which affects the accuracy of the result.
We deduce
$$
\Psi_5(0) ~\simeq~ N_F  ~ \bigl(3.70 \pm 0.90 \pm 0.30
\pm 0.70 \pm 2.00 \bigr)
10^{-6}\GV^4
,\eqno(B.21)
$$
where the errors are due to $f$, $\L$, $\langle\a_sG^2\rangle$
and $\r\a_s\langle\bar u u\rangle^2$.
Adding these errors quadratically, we obtain
$$
\sqrt{\Psi_5(0)} ~\simeq~   \sqrt{N_F}~ (1.92 \pm 0.53 )10^{-3} \GV^2
.\eqno(B.22)
$$
Using this value in eq.(B.9) (with $\tilde m_{\hp}$), after multiplying
the numerator and denominator by the overall $2m_R$ factor and using
Dashen's formula for $m_R\langle \phi_R \rangle$,
we finally find
$$
F~\simeq~ (1.55 \pm 0.43) \sqrt{2N_F}~f_\p
,\eqno(B.23)
$$
to be compared with the OZI prediction of $\sqrt{2N_F}~f_\p$.

This result is again in broad agreement with our expectations,
although of course the errors are much too large to draw
a definitive conclusion. Nevertheless, this confirmation
can be taken as providing extra support for the reliability
of the estimate in the text for $\chi^{\prime}(0)$.

\vskip1cm
\noindent{\bf Acknowledgements}
\vskip0.3cm
One of us (GMS) would like to thank John Ellis and the TH Division,
CERN for their hospitality during several visits.

\vfill\eject

\noindent{\bf Figure captions}

\vskip1cm
\noindent{\bf Fig. 1}~~~~The two-current matrix element
$\langle N|J_\m(q)~J_\n(-q)|N\rangle$.

\vskip 0.5cm
\noindent{\bf Fig. 2}~~~~Decomposition of the matrix element into a
composite operator propagator~(denoted by the double line)
and a proper vertex (hatched).

\vskip 0.5cm
\noindent{\bf Fig. 3}~~~~The original parton model representation
of the scattering amplitude.

\vskip 0.5cm
%\vskip4cm
\noindent{\bf Fig. 4}~~~~The QCD-improved parton model representation.

\vskip 0.5cm
\noindent{\bf Fig. 5}~~~~a) $\tau$-behaviour of
$\sqrt{\chi^{\prime}(0)}$ for different values of the
continuum threshold $t_c$.

\noindent \hskip1.5cm b) behaviour of different $\tau$-minima
versus $t_c$.

\vskip 0.5cm
\noindent{\bf Fig. 6}~~~~a) As Fig. 5a for the parameter
$f_{\eta_{\prime}}$.

\noindent \hskip1.5cm b) FESR prediction of $\sqrt{\chi^{\prime}(0)}$
versus $t_c$ for different values of $f_{\hp}$.

\vskip 0.5cm
\noindent{\bf Fig. 7}~~~~As Fig. 5a for the parameter $f$.

\vskip 0.5cm
\noindent{\bf Fig. 8}~~~~As Fig. 5a for $\Psi_5(0)$.

\vfill\eject

\noindent{\bf References}
\vskip0.5cm

\settabs\+\ [&1] &G.M.Shore  \cr

\+\ [&1] &J. Ashman et al., Phys. Lett. B206 (1988) 364;
Nucl. Phys. B328 (1990) 1 \cr
\+\ [&2] &G. Baum et al., Phys. Rev. Lett. 51 (1983) 1135 \cr
\+\ [&3] &G. Altarelli, Lectures at the International School
of Subnuclear Physics, Erice \cr
\+  &{}  &CERN-TH.5675/90  \cr
\+\ [&4] &S.D. Bass and A.W. Thomas, J. Phys. G19 (1993) 925 \cr
\+\ [&5] &G. Altarelli and G.G. Ross, Phys. Lett. B212 (1988) 391 \cr
\+\ [&6] &G.M. Shore and G. Veneziano, Phys. Lett. B244 (1990) 75 \cr
\+\ [&7] &G.M. Shore and G. Veneziano, Nucl. Phys. B381 (1992) 23 \cr
\+\ [&8] &R.G. Roberts, ``The Structure of the Proton'',
Cambridge University Press (1990) \cr
\+\ [&9] &J. Ellis and R.L. Jaffe, Phys. Rev. D9 (1974) 1444, D10
(1974) 1669 \cr
\+ [1&0] &J. Kodaira, Nucl. Phys. B165 (1980) 129  \cr
\+ [1&1] &S.A. Larin and J.A.M Vermaseren,
Phys. Lett. B259 (1991) 213 \cr
\+ &{} &S.A. Larin, CERN-TH.7208/94 \cr
\+ [1&2] &D. Espriu and R. Tarrach, Z. Phys. C16 (1982) 77 \cr
\+ [1&3] &G. Veneziano, Lecture at the Okubofest, May 1990,
CERN-TH.5840/90 \cr
\+ [1&4] &G. Veneziano, Mod. Phys. Lett. A4 (1989) 1605 \cr
\+ [1&5] &M. Bourquin et al, Z. Phys. C21 (1983) 27 \cr
\+ [1&6] &Z. Dziembowski and J. Franklin, J. Phys. G17 (1991) 213  \cr
\+ &{} &F.E. Close and R.G. Roberts, Phys. Lett. B302 (1993) 533 \cr
\+ [1&7] &E. Witten, Nucl. Phys. B156 (1979) 269 \cr
\+ [1&8] &G. Veneziano, Nucl. Phys. B159 (1979) 213 \cr
\+ [1&9] &M.A. Shifman, A.I. Vainshtein and V.I. Zakharov,
Nucl. Phys. B147 (1979) 385, 448 \cr
\+ [2&0] &S. Narison, QCD Spectral Sum Rules, Lecture Notes in Physics,
Vol.~26, \cr
\+&{} &World Scientific, Singapore, 1989 \cr
\+ [2&1] &G.M. Shore, Nucl. Phys. B362 (1991) 85 \cr
\+ [2&2] &A.L. Kataev, N.V. Krasnikov and A.A. Pivovarov,
Nucl. Phys. B198 (1982) 508 \cr
\+ [2&3] &V.A. Novikov et al., Nucl. Phys. B237 (1984) 525 \cr
\+ [2&4] &S. Narison, Z. Phys. C26 (1984) 209 \cr
\+ [2&5] &S. Narison, Phys. Lett. B255 (1991) 101 \cr
\+ [2&6] &Particle Data Group, Phys. Rev. D45, Part 2 (June 1992) \cr
\+ [2&7] &R.A. Bertlmann, G. Launer and E. de Rafael, Nucl. Phys.
B250 (1985) 61 \cr
\+ [2&8] &G. Briganti, A. Di Giacomo and H. Panagopoulos, Phys.
Lett. B253 (1991) 427   \cr
\+ [2&9] &A. Di Giacomo, Lecture at the QCD 90 Workshop, Montpellier,
July 1990, \cr
\+&{} & Nucl. Phys. (Proc. Suppl.) 23B (1991) \cr
\+ [3&0]&E. Braaten, A. Pich and S. Narison,
Nucl. Phys. B373 (1992) 581  \cr
\+&{} &ALEPH: D. Buskulic et al., Phys. Lett. B307 (1993) 209  \cr
\+&{} &S. Narison, Talk given at the Fubini fest  (Torino 1994),
 CERN-TH.7188/94\cr
\+&{} &and references quoted therein. \cr
\+ [3&1] &SMC: P. Shanahan, talk given at
the 29th rencontre de Moriond \cr
\+&{} &on QCD and high-energy hadronic
interactions, Meribel, Savoie, March 19-26 1994      \cr
\+&{} &D. Adams et al. CERN-PPE/94-57 (1994) (to be published)\cr
\+ [3&2] &SMC: B. Aveda et al, Phys. Lett. B302 (1993) 533  \cr
\+ [3&3] &E142: P.L. Anthony et al, Phys. Rev. Lett. 71 (1993) 959 \cr
\+ [3&4] &J.D. Bjorken, Phys. Rev. 148 (1966) 1467, D1
(1970) 1376 \cr
\+ [3&5] &SMC: B. Aveda et al, Phys. Lett. B320 (1994) 400  \cr
\+ [3&6] &I.I. Balitsky, V.M. Braun and A.V. Kolesnichenko, \cr
\+&{} &Phys. Lett. B242 (1990) 245; B318 (1993) 648 (E)\cr
\+&{} &X. Ji and M. Unrau, MIT-CTP-2232 (1993). \cr
\+ [3&7] &G.C. Ross and R.G. Roberts, Phys. Lett. B322 (1994) 425 \cr
\+ [3&8] &S. Narison, G.M. Shore and G. Veneziano,
Nucl. Phys. B391 (1993) 69 \cr
\+ [3&9] &L. Trentadue and G. Veneziano, Phys. Lett. B323 (1994) 201 \cr
\+ [4&0] &S.D. Bass, Z. Phys. C55 (1992) 653 \cr
\+ [4&1] &G.M. Shore and G. Veneziano, Nucl. Phys. B381 (1992) 3 \cr
\+ [4&2] &M.S. Chanowitz, {\it in} Proc. VI Int. Workshop on
Photon-Photon Collisions, \cr
\+&{}&Lake Tahoe, 1984, ed. R. Lander, World Scientific,
Singapore, 1984 \cr
\+ [4&3] &S. Narison, Phys. Lett. B104 (1981) 485  \cr

\vfill\eject

\bye